\documentclass[conference]{IEEEtran}
\IEEEoverridecommandlockouts
\usepackage{cite}
\usepackage{amsmath,amssymb,amsfonts}
\usepackage{algorithmic}
\usepackage{graphicx}
\usepackage{textcomp}
\usepackage{xcolor}
\usepackage[utf8]{inputenc}
\usepackage{listings}
\usepackage{enumitem}
\usepackage{caption}

\newcounter{listcounter}
\renewcommand{\thelistcounter}{\arabic{listcounter}}

\newcommand{\jecaption}[2]{%
    \vspace{-0.5em} 
    \refstepcounter{listcounter} 
    \begin{center}
    \footnotesize List \thelistcounter: #1
    \label{#2}
    \end{center}
}

\def\BibTeX{{\rm B\kern-.05em{\sc i\kern-.025em b}\kern-.08em
    T\kern-.1667em\lower.7ex\hbox{E}\kern-.125emX}}
    
\begin{document}

\title{Proposal of an Automatic Verification Method for Network Configuration Model by Static Analysis}

\author{
    \IEEEauthorblockN{Tomoya Fujita}
    \IEEEauthorblockA{
        \textit{Shinshu University}\\
        Nagano, Japan \\
        24w6063b@shinshu-u.ac.jp
    }
    \and
    \IEEEauthorblockN{Hikofumi Suzuki}
    \IEEEauthorblockA{
        \textit{National Institute of Informatics}\\
        Tokyo, Japan \\
        h-suzuki@nii.ac.jp
    }
    \and
    \IEEEauthorblockN{Shinpei Ogata}
    \IEEEauthorblockA{
        \textit{Shinshu University}\\
        Nagano, Japan \\
        ogata@cs.shinshu-u.ac.jp
    }
    \and
    \IEEEauthorblockN{Hiroaki Hashiura}
    \IEEEauthorblockA{
        \textit{Nippon Institute of Technology}\\
        Saitama, Japan \\
        hashiura@nit.ac.jp
    }
    \and
    \IEEEauthorblockN{Takashi Nagai}
    \IEEEauthorblockA{
        \textit{Institute of Technologists}\\
        Saitama, Japan \\
        t\_nagai@iot.ac.jp
    }
    \and
    \IEEEauthorblockN{Kozo Okano}
    \IEEEauthorblockA{
        \textit{Shinshu University}\\
        Nagano, Japan \\
        okano@cs.shinshu-u.ac.jp
    }
}

\maketitle

\begin{abstract}
In the network design phase, designers typically assess the validity of the network configuration on paper. However, the interactions between devices based on network protocols can be complex, making this assessment challenging. Meanwhile, testing with actual devices incurs significant costs and effort for procurement and preparation. Traditional methods, however, have limitations in identifying configuration values that cause policy violations and verifying syntactically incomplete device configuration files. In this paper, we propose a method to automatically verify the consistency of a model representing the network configuration (Network Configuration Model) by static analysis. The proposed method performs verification based on the network configuration model to detect policy violations and points out configuration values that cause these violations. Additionally, to facilitate the designers' review of each network device’s configuration, the model is converted into a format that mimics the output of actual devices, which designers are likely familiar with. As a case study, we applied the proposed method to the network configuration of Shinshu University, a large-scale campus network, by intentionally introducing configuration errors and applying the method. We further evaluated whether it could output device states equivalent to those of actual devices.
\end{abstract}


\section{Introduction}
Networks are one of the indispensable IT infrastructures in numerous organizational activities\cite{downtime}, and failures in their design are becoming increasingly unacceptable.
In general, the validity of a network design is verified through construction and testing on actual network devices\cite{miyata_siken}.
However, testing with actual devices requires significant cost and effort for procurement and preparation.
When actual devices are not available, network engineers manually validate the design on paper based on their own expertise\cite{kijokensa_tajima}.
Since the interactions between devices based on network protocols are complex\cite{gember2016fast}, such paper-based validation becomes impractical, especially for large-scale and complex networks.
Therefore, as a method to support design validation without relying on actual devices, previous studies have focused on verifying network designs using only device configurations and network topology\cite{kijokensa_tajima,batfish}.
In the study by Tajima et al.\cite{kijokensa_tajima}, the consistency of device configurations can be verified by analyzing topology data abstracted into models.
Similarly, Batfish\cite{batfish} simulates the behavior of all protocols to generate forwarding tables and subsequently uses dataplane verification tools to validate network properties.

One example of the design aspects to be verified is adherence to network operation policies, such as ensuring that there is no IP address duplication and that VLAN settings are consistent between peer devices.
However, conventional methods\cite{kijokensa_tajima,batfish} cannot directly identify the configuration values that cause policy violations, leaving engineers with the burden of manually pinpointing them.
Moreover, device configurations like those used as input by Batfish suffer from the problem that a single syntax error can render the entire configuration unparseable.
This issue arises because engineers manually create configuration files as single textual documents, even though these files are based on a defined grammar.

To address such problems, we have previously proposed a network configuration model\cite{arai,nakamura}.
The network configuration model represents both the network structure and device configurations by applying the object-oriented modeling language UML (Unified Modeling Language).
A key feature of this model is that it expresses the necessary configuration items in a structured diagrammatic form.
Thus, engineers only need to provide configuration values for each item, without having to construct textual sentences that concatenate different items.
This feature ensures that even if a single configuration value contains an error, the entire model does not become unparseable, allowing direct identification of only the problematic configuration values.
However, a remaining challenge was that no method had yet been established to automatically validate network operation policies based on this model.

Therefore, this study aims to develop an automated verification method that statically analyzes network configuration model to detect policy violations and identify the specific configuration values responsible for them.

In the proposed method, the network configuration model is verified to detect policy violations and identify the specific configuration parameters responsible for them. Furthermore, to reduce the effort required for designers to extract information from the model, its contents are converted into a format consistent with the output of actual network devices.

For evaluation, the proposed method was applied to an operational network running the OSPF (Open Shortest Path First) protocol, with intentional configuration errors introduced. The results demonstrated that the method detected 90\% of policy violations and accurately identified the configuration values responsible for them in 80\% of the cases.
After improvements to the proposed method, it was able to detect policy violations that had previously gone undetected, demonstrating its potential effectiveness.
Furthermore, we evaluated whether the output of the model transformation was equivalent to the state information obtained from actual network devices.
The results confirmed equivalence for all information except for the states that depend on the startup order of the devices.

The proposed method is an improvement over our previous work \cite{fujita_iot}.
While the previous method could not handle the OSPF protocol, the proposed method extends support to it.
Moreover, whereas the previous study \cite{fujita_iot} evaluated the method on a constructed test network, in this study we evaluated the proposed method on an operational network.

The remainder of this paper is organized as follows.
Section 2 provides an overview of the technologies addressed in this study.
Section 3 describes the proposed method.
Section 4 presents the evaluation experiments and their results, demonstrating the effectiveness of the proposed method.
Section 5 discusses related work, and Section 6 concludes the paper and outlines directions for future research.
\begin{figure*}[!t]
  \centering
  \includegraphics[width=\linewidth]{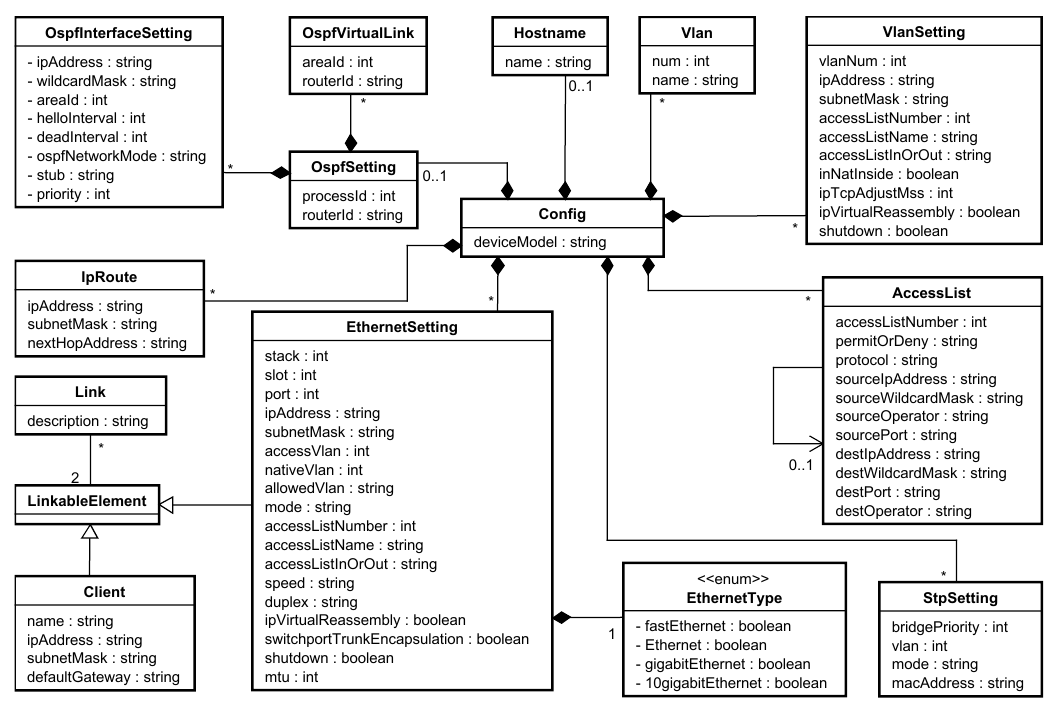}
  \caption{Network configuration metamodel}
  \label{fig:networkkouseimetamodel}
\end{figure*}

\begin{figure}[!t]
  \centering
  \includegraphics[width=0.9\columnwidth]{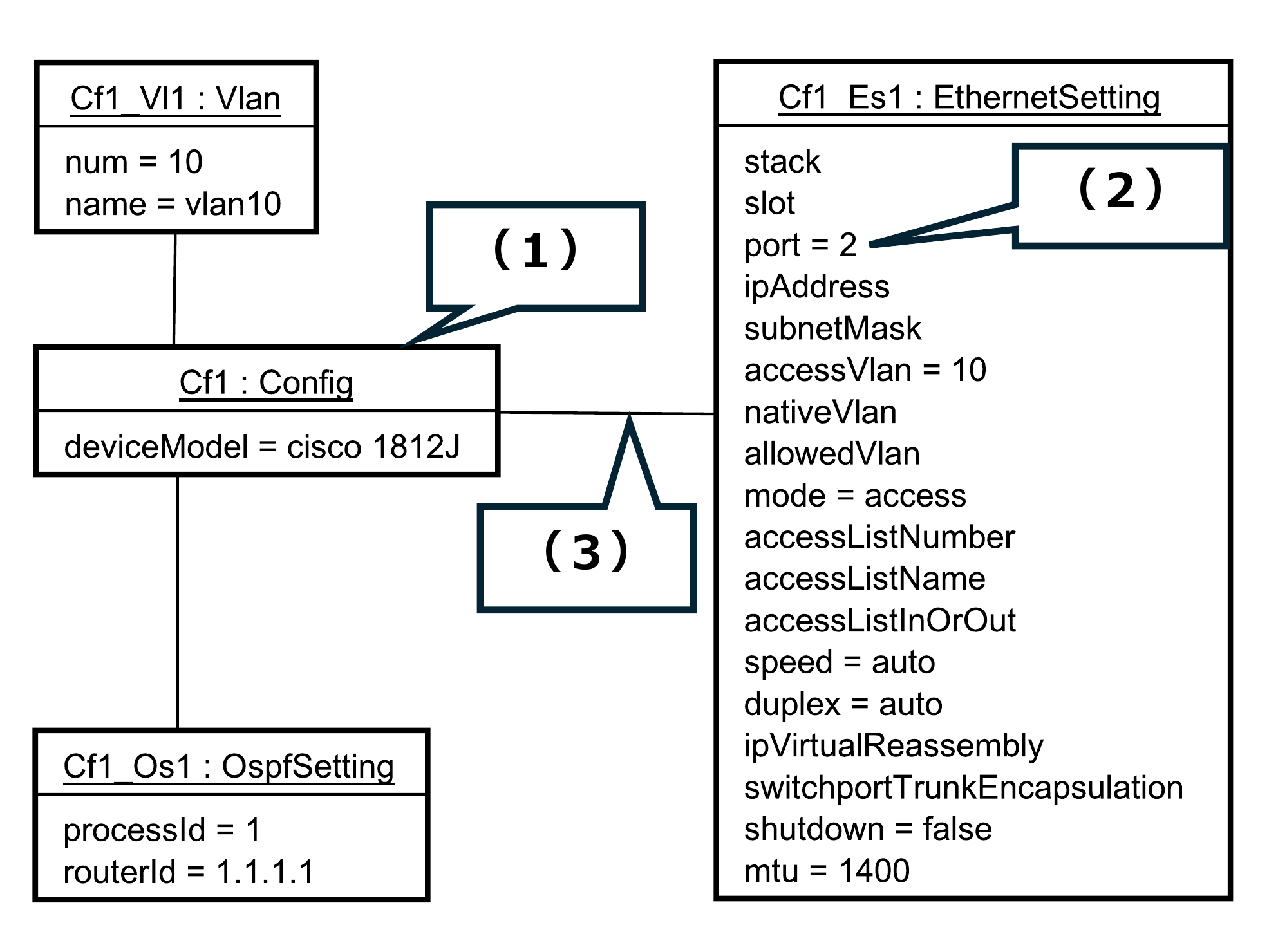}
  \caption{Network configuration model}
  \label{fig:networkkouseimodel}
\end{figure}

\section{Preliminaries}
In this chapter, we review prior research and the technologies employed in the proposed method.
\subsection{Network Configuration Model}
The network configuration model\cite{arai,nakamura} is a representation of network topology and device configurations using UML object diagram notation.
The network configuration model contains a network configuration metamodel, in which specification items are structurally defined using UML class diagram notation.
Fig.~\ref{fig:networkkouseimetamodel} shows the network configuration metamodel.
As the main structural concept, there are \textbf{specification item groups} (e.g., a large rectangle labeled VlanSetting) that consist of zero or more closely related \textbf{specification items} (e.g., num:int).
These specification items are primarily defined based on the command hierarchy of Cisco network devices \cite{arai}.
For example, an OSPF network configuration (e.g., network 172.16.1.0 0.0.0.255 area 0) is represented in the model using OspfInterfaceSetting.
In this example, "172.16.1.0" represents the IP address and is expressed as the ipAddress configuration value, "0.0.0.255" represents the wildcard mask and is expressed as the wildcardMask configuration value, and "0" represents the area and is expressed as the areaId configuration value.

In addition, there are \textbf{relationships} between specification item groups (e.g., the line between Config and VlanSetting).

Fig.~\ref{fig:networkkouseimodel} shows the network configuration model.
This model is designed to satisfy the following three conditions.

(1) Create values (hereafter referred to as nodes) typed by the specification item groups of the metamodel (e.g., a large rectangle labeled Cf1:Config). Here, Config is referred to as the specification item group name, and Cf1 as the specification item group value name.

(2) Assign values (hereafter referred to as configuration values) to the specification items in each group (e.g., the 2 in port = 2).

(3) Describe the lines (hereafter referred to as association lines) between group values in a way that is consistent with the relationships among the specification item groups.

The metamodel enables both physical design aspects, such as connections between network devices, and logical design aspects, such as individual device configurations (e.g., EthernetSetting or Vlan), to be described consistently using a single notation.
Due to these characteristics, verification at the level of individual configuration values is always feasible, and by following the association lines, the consistency of configurations across devices can be validated.
Therefore, in this study, the network configuration model is adopted as the specification for network configurations.

\subsection{Static Analysis}
Static analysis \cite{batfish} refers to a method that directly analyzes the configuration files of network devices.
One advantage of this approach is that it enables errors to be detected before new configurations are actually applied to the network.
For example, rcc \cite{rcc} can detect issues such as duplicate router IDs or missing configurations in BGP (Border Gateway Protocol).
Since network design and validation are often performed manually, even simple mistakes may be overlooked when designing large-scale networks by hand, potentially leading to network malfunctions \cite{arai1}.
In this study, we adopt a static analysis approach to detect policy violations, thereby enabling verification of the design quality of network configurations without requiring actual devices.
\begin{figure*}[!t]
  \centering
  \includegraphics[width=0.8\linewidth]{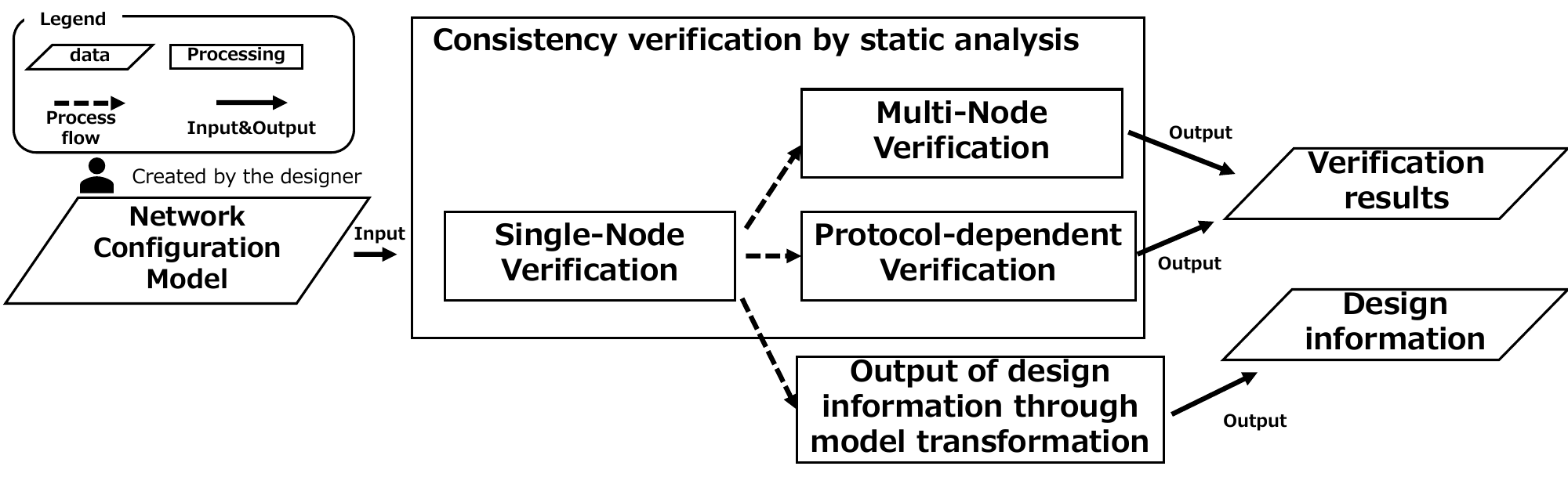}
  \caption{Overview of the proposed method}
  \label{fig:teiansyuhou_zentaizu}
\end{figure*}

\section{Proposed Method}
Fig.~\ref{fig:teiansyuhou_zentaizu} shows an overview of the proposed method.
The proposed method assumes a situation in which a network designer (hereafter referred to as the designer) describes new configurations to be applied to a network within a network configuration model (hereafter referred to as the model).
By using the proposed method, the designer can verify the consistency of configurations (\ref{ikkansei}) and check them in the format output by actual network devices (\ref{syuturyoku}).

\subsection{Consistency Verification by Static Analysis}\label{ikkansei}
The purpose of consistency verification is to identify configuration values that violate lexical rules and to detect inconsistencies among configurations..
First, single-node verification (\ref{tantai}) is performed to detect invalid configuration values and missing mandatory configuration values.
If no errors are found in the single-node verification, multi-node verification (\ref{fukusu}) and protocol-dependent verification (\ref{protocol}) are subsequently performed.
Based on the results of these verifications, the designer refines the network by modifying the model.

\subsubsection{Single-Node Verification}\label{tantai}
In single-node verification, we verify whether the configuration values extracted from the model are syntactically valid for network devices and whether they are consistent with other configuration values within the same node.
Of the 71 configuration values defined in the network configuration metamodel, 68 used in device configurations have defined lexical and syntactic rules.

In the proposed method, to help designers better understand the causes of verification errors, lexical and syntactic rules are organized based on the classification shown in TABLE~\ref{tab:bunrui}.
The notation “Regular Expression satisfying Criterion” follows the syntax of Java 17 regular expressions.
In the type-based classification, the corresponding regular expressions for the int and boolean types, along with the outputs when a violation occurs, are shown.
For key and format, because the regular expressions and violation outputs vary according to the configuration value, examples are provided for the accessVlan and mode configuration values of the EthernetSetting node.
TABLE~\ref{tab:check} shows the relationship between specification items and lexical and syntactic rules.
In the table, the symbol “$ \square $” denotes the presence of a rule, whereas “–” denotes its absence.

To verify consistency with other configuration values within the same node, the contents of TABLE~\ref{tab:int_verif} are checked.
In Case (a), when an EthernetSetting node exists and either ipAddress or subnetMask is set, the other is also set. Case (b) is checked similarly.
Cases (c)–(f) verify that when a specified node (e.g., EthernetSetting) exists, mandatory configuration values (e.g., port) are present.
If any mandatory values are missing, an error message is displayed.

\begin{table*}[t]
\centering
\caption{Overview of lexical and syntactic rules}
\label{tab:bunrui}
\renewcommand{\arraystretch}{1.2}
\scalebox{0.88}{
\begin{tabular}{lllll}
\hline\hline
Category& Description& Regular Expression satisfying Criterion  &Violation output \\ 
\hline 
Non-empty & No spaces allowed within the value &{\ttfamily \^{}\textbackslash S+\$} &Input contains spaces\\ 
Non-full-width & No full-width characters allowed within the value &{\ttfamily \^{}[\textbackslash x00-\textbackslash x7F]*\$} &Input contains full-width characters\\ 
Type & Value must match the specified type (int, boolean)  & {\ttfamily \^{}-?\textbackslash d+\$}, {\ttfamily \^{}(true|false)\$}& Enter an integer / enter a boolean\\ 
Key & Must use a keyword as the value &{\ttfamily \^{}(access|trunk)\$}& enter either 'access' or 'trunk'\\
Format & Value must follow a specific string format &  {\ttfamily \^{}(40[0-8][0-9]|409[0-4]|[1-9][0-9]\{0,2\})\$} & Enter an integer between 1 and 4094\\ \hline
\end{tabular}
}
\end{table*}

\begin{table}[t]
\centering
\caption{The relationship between specification item and lexical or syntactic rules}
\label{tab:check}
\renewcommand{\arraystretch}{1.13}
\scalebox{0.65}{
\begin{tabular}{llcp{1.2cm}ccc}
\hline\hline
 specification item group &  specification item & Non-empty & Non-full-width & Type & Key & Format \\ \hline
HostName  & name  & $ \square $ & $ \square $ & ‐ & ‐ & ‐ \\ \hline
EthernetSetting & stack & $ \square $ & $ \square $ & $ \square $ & ‐ & $ \square $ \\ \cline{2-7}
 & slot & $ \square $ & $ \square $ & $ \square $ & ‐ & $ \square $ \\ \cline{2-7}
 & port & $ \square $ & $ \square $ & $ \square $ & ‐ & $ \square $ \\ \cline{2-7}
 & ipAddress & $ \square $ & $ \square $ & ‐ & ‐ & $ \square $ \\ \cline{2-7}
 & subnetMask & $ \square $ & $ \square $ & ‐ & ‐ & $ \square $ \\ \cline{2-7}
 & accessVlan & $ \square $ & $ \square $ & $ \square $ & ‐ & $ \square $ \\ \cline{2-7}
 & nativeVlan & $ \square $ & $ \square $ & $ \square $ & ‐ & $ \square $ \\ \cline{2-7}
 & allowedVlan & $ \square $ & $ \square $ & ‐ & ‐ & $ \square $ \\ \cline{2-7}
 & mode & $ \square $ & $ \square $ & ‐ & $ \square $ & ‐ \\ \cline{2-7}
 & accessListNumber & $ \square $ & $ \square $ & $ \square $ & ‐ & $ \square $ \\ \cline{2-7}
 & accessListName & $ \square $ & $ \square $ & ‐ & ‐ & ‐ \\ \cline{2-7}
& accessListInOrOut & $ \square $ & $ \square $ & ‐ & $ \square $ & ‐ \\ \cline{2-7}
 & speed & $ \square $ & $ \square $ & ‐ & $ \square $ & ‐ \\ \cline{2-7}
 & duplex & $ \square $ & $ \square $ & ‐ & $ \square $ & ‐ \\ \cline{2-7}
 & ipVirtualReassembly & $ \square $ & $ \square $ & $ \square $ & ‐ & ‐ \\ \cline{2-7}
 & switchportTrunkEncapsulation & $ \square $ & $ \square $ & $ \square $ & ‐ & ‐ \\ \cline{2-7}
 & shutdown & $ \square $ & $ \square $ & $ \square $ & ‐ & ‐ \\ \cline{2-7}
 & mtu &$ \square $ & $ \square $ & $ \square $ & ‐ & $ \square $\\ \hline
Vlan & num & $ \square $ & $ \square $ & $ \square $ & ‐ & $ \square $ \\ \cline{2-7}
 & name & $ \square $ & $ \square $ & ‐ & ‐ & ‐ \\ \hline
VlanSetting & vlanNum & $ \square $ & $ \square $ & $ \square $ & ‐ & $ \square $ \\ \cline{2-7}
 & ipAddress & $ \square $ & $ \square $ & ‐ & ‐ & $ \square $ \\ \cline{2-7}
 & subnetMask & $ \square $ & $ \square $ & ‐ & ‐ & $ \square $ \\ \cline{2-7}
 & accessListNumber & $ \square $ & $ \square $ & $ \square $ & ‐ & $ \square $ \\ \cline{2-7}
 & accessListName & $ \square $ & $ \square $ & ‐ & ‐ & ‐ \\ \cline{2-7}
 & accessListInOrOut & $ \square $ & $ \square $ & ‐ & ‐ & ‐ \\ \cline{2-7}
 & ipNatInside & $ \square $ & $ \square $ & $ \square $ & ‐ & $ \square $ \\ \cline{2-7}
 & ipTcpAdjustMss & $ \square $ & $ \square $ & $ \square $ & ‐ & ‐ \\ \cline{2-7}
 & ipVirtualReassembly & $ \square $ & $ \square $ & $ \square $ & ‐ & ‐ \\ \cline{2-7}
 & shutdown & $ \square $ & $ \square $ & $ \square $ & ‐ & ‐ \\ \hline
StpSetting & bridgePriority & $ \square $ & $ \square $ & $ \square $ & ‐ & $ \square $ \\ \cline{2-7}
 & Vlan & $ \square $ & $ \square $ & $ \square $ & ‐ & $ \square $ \\ \cline{2-7}
 & Mode & $ \square $ & $ \square $ & ‐ & $ \square $ & ‐ \\ \cline{2-7}
 & macAddress & $ \square $ & $ \square $ & ‐ & ‐ & $ \square $ \\ \hline
IpRoute & Network & $ \square $ & $ \square $ & ‐ & ‐ & $ \square $ \\ \cline{2-7}
 & addressPrefix & $ \square $ & $ \square $ & ‐ & ‐ & $ \square $ \\ \cline{2-7}
 & nextHopAddress & $ \square $ & $ \square $ & ‐ & ‐ & $ \square $ \\ \hline
OspfVirtualLink & areaId & $ \square $ & $ \square $ & $ \square $ & ‐ & $ \square $ \\ \cline{2-7}
 & routerId & $ \square $ & $ \square $ & ‐ & ‐ & $ \square $ \\ \hline
OspfInterfaceSetting & ipAddress & $ \square $ & $ \square $ & ‐ & ‐ & $ \square $ \\ \cline{2-7}
 & wildcardMask & $ \square $ & $ \square $ & ‐ & ‐ & $ \square $ \\ \cline{2-7}
 & areaId & $ \square $ & $ \square $ & $ \square $ & ‐ & $ \square $ \\ \cline{2-7}
 & helloInterval & $ \square $ & $ \square $ & $ \square $ & ‐ & $ \square $ \\ \cline{2-7}
 & deadInterval & $ \square $ & $ \square $ & $ \square $ & ‐ & $ \square $ \\ \cline{2-7}
 & ospdNetworkMode & $ \square $ & $ \square $ & ‐ & $ \square $ & ‐ \\ \cline{2-7}
 & stub & $ \square $ & $ \square $ & ‐ & $ \square $ & ‐ \\ \cline{2-7}
& priority & $ \square $ & $ \square $ & $ \square $ & ‐ & $ \square $ \\ \hline 
OspfSetting & processId & $ \square $ & $ \square $ & $ \square $ & ‐ & ‐ \\ \cline{2-7}
 & routerId & $ \square $ & $ \square $ & ‐ & ‐ & $ \square $ \\ \hline
Client & IpAddress & $ \square $ & $ \square $ & ‐ & ‐ & $ \square $ \\ \cline{2-7}
 & subnetMask & $ \square $ & $ \square $ & $ \square $ & ‐ & $ \square $ \\ \cline{2-7}
 & defaultGateway & $ \square $ & $ \square $ & ‐ & ‐ & $ \square $ \\ \hline
EthernetType & fastEthernet & $ \square $ & $ \square $ & $ \square $ & ‐ & ‐ \\ \cline{2-7}
 & Ethernet & $ \square $ & $ \square $ & $ \square $ & ‐ & ‐ \\ \cline{2-7}
 & gigabitEthernet & $ \square $ & $ \square $ & $ \square $ & ‐ & ‐ \\ \cline{2-7}
 & 10gigabitEthernet & $ \square $ & $ \square $ & $ \square $ & ‐ & ‐ \\ \hline
AccessList & accessListNumber & $ \square $ & $ \square $ & $ \square $ & ‐ & $ \square $ \\ \cline{2-7}
 & permitOrDeny & $ \square $ & $ \square $ & ‐ & $ \square $ & ‐ \\ \cline{2-7}
 & protocol & $ \square $ & $ \square $ & ‐ & $ \square $ & ‐ \\ \cline{2-7}
 & sourceIpAddress & $ \square $ & $ \square $ & ‐ & ‐ & $ \square $ \\ \cline{2-7}
 & sourceWildcardMask & $ \square $ & $ \square $ & ‐ & ‐ & $ \square $ \\ \cline{2-7}
 & sourceOperator & $ \square $ & $ \square $ & ‐ & $ \square $ & ‐ \\ \cline{2-7}
 & sourcePort & $ \square $ & $ \square $ & ‐ & ‐ & $ \square $ \\ \cline{2-7}
 & destIpAddress & $ \square $ & $ \square $ & ‐ & ‐ & $ \square $ \\ \cline{2-7}
 & destWildcardMask & $ \square $ & $ \square $ & ‐ & ‐ & $ \square $ \\ \cline{2-7}
 & destPort & $ \square $ & $ \square $ & ‐ & ‐ & $ \square $ \\ \cline{2-7}
 & destOperator & $ \square $ & $ \square $ & ‐ & $ \square $ & ‐ \\ \hline
 \end{tabular}
 }
\end{table}

\begin{table}[t]
\centering
\caption{Verification items related to configuration values within a node}
\label{tab:int_verif}
\renewcommand{\arraystretch}{1.2}
\scalebox{0.95}{
\begin{tabular}{ll}
\hline\hline
Verification items \\ \hline 
(a)Inconsistency between ipAddress and subnetMask in EthernetSetting\\
(b)Inconsistency  between mode accessVlan, and nativeVlan in VlanSetting \\
(c)Missing port in EthernetSetting \\
(d)Missing ipAddress, wildcardMask, or areaId in OspfInterfaceSetting\\ 
(e)Missing vlanNum in VlanSetting \\ 
(f)Missing num in Vlan\\ \hline
\end{tabular}
}
\end{table}

\subsubsection{Multi-Node Verification}\label{fukusu}
In multi-node verification, we detect configuration errors that may lead to abnormal behavior due to interactions between devices, such as inconsistencies between connected network devices or duplicate IP addresses within the network.
Because the model represents device physical connections using Link nodes and includes detailed configurations such as IP addresses and VLAN IDs, it enables verification across Layers 1, 2, and 3.
TABLE~\ref{tab:fukusu} lists the verification items for multiple nodes.
This paper presents the verification methods for two of these items.

First, we explain the verification method for “allowedVlan mismatch” in TABLE~\ref{tab:fukusu}.
allowedVlan refers to the VLANs permitted on a trunk port.
If allowedVlan does not match between peer devices, this may indicate an error in the allowedVlan configuration of the EthernetSetting node or the omission of a Vlan node.
Therefore, such inconsistencies are detected.
The procedure for verifying “allowedVlan mismatch” in the model is as follows:

\begin{enumerate}
\item For all pairs of EthernetSetting nodes $(e_1, e_2)$ connected via a Link node, proceed to the next steps if the mode configuration values of both $e_1$ and $e_2$ are set to “trunk.” Steps (2)–(4) are applied to both $e_1$ and $e_2$.
\item Obtain the set of VLAN numbers $M$ from the allowedVlan configuration values of the EthernetSetting node.
\item Retrieve the Config node connected via an association line to the EthernetSetting node, and then obtain the num configuration values of all Vlan nodes connected to it. Define this set of VLAN numbers as $N$.
\item Store the result of $M \cap N$ in a list.
\item If the lists obtained from $e_1$ and $e_2$ do not match, report an “inconsistency of allowedVlan” for both $e_1$ and $e_2$.
\end{enumerate}

\begin{table*}[ht]
\centering
\caption{Verification items for validation across multiple nodes}
\label{tab:fukusu}
\renewcommand{\arraystretch}{1.2}
\scalebox{0.97}{
\begin{tabular}{lll}
\hline\hline
items                     &  Problem                                  \\ \hline
(1)Duplicate IP addresses on the same device        &  Error occurs when issuing configuration commands\\
(2)Duplicate IP addresses on different devices & Communication may be disrupted if duplicate IP addresses exist within the same segment\\
(3)VLAN duplication             & Unintended VLAN configurations may have been applied \\ 
(4)allowedVlan mismatch  &  The interface may allow incorrect VLANs or some VLANs may not have been created               \\ 
(5)accessVlan mismatch  &   An incorrect VLAN may have been configured on the port \\
(6)nativeVla mismatchn   & Communication failure with the peer device \\
(7)duplex mismatch & Unstable communication caused by packet collisions\\
(8)speed mismatch & Possible unstable communication\\
(9)Loop  & If not addressed, a broadcast storm may increase network load and disrupt communication\\
(10)Overlapping subnets across different segments &Packets may be misdelivered to other devices, leading to unstable communication\\
(11)Assignment of different networks within the same VLAN & Possible misconfiguration of VLAN or IP address assignment \\
(12)Release of unused interfaces \footnotemark[1] & If an interface is unintentionally accessible from outside, unexpected behavior may occur \\
\hline
\end{tabular}
}
\end{table*}
\footnotetext[1]{Items added through the experiments in this paper}

Next, we describe the verification method for “Duplicate IP addresses on different devices.”
If IP addresses are duplicated within the same segment, this may cause packet loss or degradation of network performance.
When IP addresses are duplicated across different segments, problems may arise when communication between segments (such as inter-VLAN communication) is performed. Although this issue can be handled by changing IP addresses or using NAT (Network Address Translation), it should be verified to prevent oversights by designers.
The proposed method detects not only IP address duplication within EthernetSetting, VlanSetting, Client, and OspfInterfaceSetting, but also duplication across different specification-item groups.
The following explains the verification procedure for “IP address duplication” when IP addresses are configured using an SVI (Switch Virtual Interface) in the model—namely, when the ipAddress of a VlanSetting is defined.

\begin{enumerate}
\item Retrieve all VlanSetting nodes and collect all their ipAddress configuration values into a list $L$. 
\item If an IP address $p$ appears more than once in $L$, obtain the set $D_1$ of VlanSetting nodes whose ipAddress equals $p$.
\item If all vlanNum values $i$ of nodes in $D_1$ are identical, proceed to step (4). Otherwise, flag all nodes in $D_1$ for “IP address duplication across different segments.”
\item For each node in $D_1$, retrieve the Config nodes connected via association links and collect them as the set $C_1$.
Then, for each node in $C_1$, retrieve the EthernetSetting nodes connected via association links, and identify any Config node $c$ whose EthernetSetting has VLAN number $i$.
The VLAN number is obtained from accessVlan when the mode is access, and from allowedVlan when the mode is trunk.
\item  Starting from $c$, perform graph traversal over the network configuration model under the following two conditions to obtain the set $C_2$ of all reachable Config nodes, including $c$:
\begin{description}
\item [Condition 1] Only Config, EthernetSetting, and Link nodes may be traversed.
\item [Condition 2] A path EthernetSetting → Link → EthernetSetting may be traversed only when the two EthernetSetting nodes connected to the Link have the same VLAN number.
\end{description}
\item For each node in $C_2$, retrieve the VlanSetting nodes connected via association links and collect them as the set $D_2$.
If the intersection $D_3 = D_1 \cap D_2$ contains two or more elements, flag all VlanSetting nodes in $D_3$ for “IP address duplication within the same segment.”
Furthermore, if the difference $D_1 \setminus D_2$ contains one or more elements, flag all VlanSetting nodes in $D_1$ for “IP address duplication across different segments.”
\end{enumerate}

\subsubsection{Protocol-dependent Verification}\label{protocol}
Protocol-dependent verification evaluates whether device configurations are consistent with the requirements of the protocols in use.
The proposed method offers verification procedures for both OSPF and STP (Spanning Tree Protocol).
When OSPF is employed, the designer must specify configurations such as area IDs and the interfaces on which OSPF is enabled.
If inconsistencies exist between the OSPF configurations of adjacent devices, OSPF adjacencies cannot be established, which leads to incorrect packet forwarding.
Accordingly, the proposed method identifies missing or inconsistent OSPF-related configurations.

The OSPF-dependent verification items are presented in TABLE~\ref{tab:ospf}.
The following subsection explains the procedure for detecting ``Area ID mismatch.''

\begin{enumerate}
\item For each pair of EthernetSetting nodes $(g_1, g_2)$ connected through a Link node, identify the corresponding OspfInterfaceSetting nodes based on their ipAddress and subnetMask configuration values. Let the node corresponding to $g_1$ be $o_1$ and the node corresponding to $g_2$ be $o_2$.
\item Compare the area IDs (areaId configuration values) of $o_1$ and $o_2$. If they differ, report an “Area ID mismatch for both $o_1$ and $o_2$.
\end{enumerate}

\begin{table}[!t]
\centering
\caption{Verification items dependent on OSPF}
\label{tab:ospf}
\renewcommand{\arraystretch}{1.2}
\begin{tabular}{ll}
\hline\hline
Items                                                      \\ \hline
(1)Missing OSPF on required interfaces       \\
(2)Detection of Duplicate Router IDs       \\
(3)Area ID mismatch\\
(4)Subnet Mismatch Between Interfaces\\
(5)HELLO interval mismatch \\
(6)DEAD interval mismatch             \\ 
(7)Area assignment to unset network addresses\\
(8)Detection of areas not connected to Area 0\\
(9)Missing required OspfInterfaceSettings\\
(10)MTU(Maximum Transmission Unit) mismatch\\
(11)Insufficient Virtual Links\footnotemark[1]\\
 (12) Missing Area 0\footnotemark[1]\\
\hline
\end{tabular} 
\end{table}

\begin{table*}[t]
\centering
\caption{Output content as design information}
\label{johosyuturyoku}
\renewcommand{\arraystretch}{1.2}
\begin{tabular}{ll}
\hline \hline
Items     & Corresponding command        \\ \hline 
VLAN information of each switch     & show vlan brief 	\textless{}Config node name	\textgreater{}      \\ 
VLAN information of all switches   &  show vlan brief all                    \\ 
Configuration information belonging to the specified VLAN & show vlan \textless{}VLANID\textgreater{}                     \\ 
Configuration file of the switch & show running config \textless{}Config node name\textgreater{}   \\ 
Spanning Tree information & show spanning-tree \textless{}Config node name\textgreater{}  \\ 
Routing protocol status & show ip protocols \textless{}Config node name\textgreater{} \\
OSPF neighbor list & show ip ospf neighbor \textless{}Config node name\textgreater{} \\
OSPF operational status per interface & show ip ospf interface \textless{}Config node name\textgreater{} \textless{}EthernetSetting node name \textgreater{} \\
OSPF operational status per interface & show ip ospf interface \textless{}Config node name\textgreater{} \textless{} VlanSetting node name\textgreater{} \\\hline
\end{tabular}
\end{table*}

\begin{figure*}[t]
    \centering
    \begin{lstlisting}[frame=lines, numbers=left]
vlan <VlanSetting:vlanNum>
  Internet Address <VlanSetting:ipAddress>/<VlanSetting:subnetMask> Area <OspfInterfaceSetting:areaId>
          Process ID <OspfSetting:processId>, Router ID <OspfSetting:routerId>,
  Network Type <OspfInterfaceSetting:ospfNetworkMode>
                          State <DR or BDR>, Priority <OspfInterfaceSetting:priority>
  Designated Router (ID) <OspfSetting:routerId>, Interface address <VlanSetting:ipAddress>
  Backup Designated router (ID) <OspfSetting:routerId>, Interface address <VlanSetting:ipAddress>
  Timer intervals configured, Hello <OspfInterfaceSetting:helloInterval>, 
  Dead <OspfInterfaceSetting:deadInterval> 
    \end{lstlisting}
    \jecaption{Output template for OSPF operational status per interface}{lst_template}
    
\end{figure*}

\subsection{Output of Design Information}\label{syuturyoku}
In the output of design information, the proposed method transforms the model and produces output in a format that emulates the output of network devices.
In the model, multiple nodes are associated with a single network device, which causes the model to become increasingly complex as the number of devices grows.
Consequently, verifying device configurations directly from the model becomes time-consuming.
To address this, the proposed method provides nine state-check commands—including device configuration files, VLAN assignments, and OSPF information—as shown in TABLE~\ref{johosyuturyoku}.
These outputs are formatted to match the style of actual network devices, allowing engineers to conveniently review and aggregate the required information.
The output format specifically emulates that of a Cisco 1812J device.

In the proposed method, the output format of actual devices during command execution is defined using templates, into which values extracted from the model are inserted.
The output template for per-interface OSPF operation status is shown in List \ref{lst_template}, and an example output with embedded values is presented in List \ref{lst_turu}.
For example, \verb|<|VlanSetting:vlanNum\verb|>| (line 1 of List \ref{lst_template}) is replaced with the vlanNum value of the specified VlanSetting node (20 in List \ref{lst_turu}).
Additionally, the operational state of each device, derived from model configuration values, is also embedded.
For example, in OSPF, the states of interfaces within each L2 segment—including DR and BDR—are determined and reported as interface status.
The DR is responsible for managing communication paths within the L2 segment, whereas the BDR acts as a backup to the DR.
For each L2 segment, one DR and one BDR are selected based on the OSPF priority values and router IDs of the devices.
In the proposed method, the DR or BDR status of each interface is determined from the OSPF priority values and router IDs in the model, and the resulting value is embedded into line 5 of List \ref{lst_template} (\verb|<|DR or BDR\verb|>|).
Some information, such as temporal details (e.g., elapsed time until the next HELLO packet), cannot be represented in the network configuration model, which constitutes a limitation.
However, since these are not related to device configurations, they do not pose any issues from the perspective of verifying device configurations.

Conventionally, verifying VLAN segments requires obtaining information from each device individually and then aggregating it.
To reduce the effort involved in this manual aggregation while following traditional verification practices, the proposed method consolidates VLAN information for the entire network from the model, which contains configuration data and other details for multiple devices, and outputs it in an aggregated form.

\begin{figure*}[!t]
  \centering
  \includegraphics[width=0.9\linewidth]{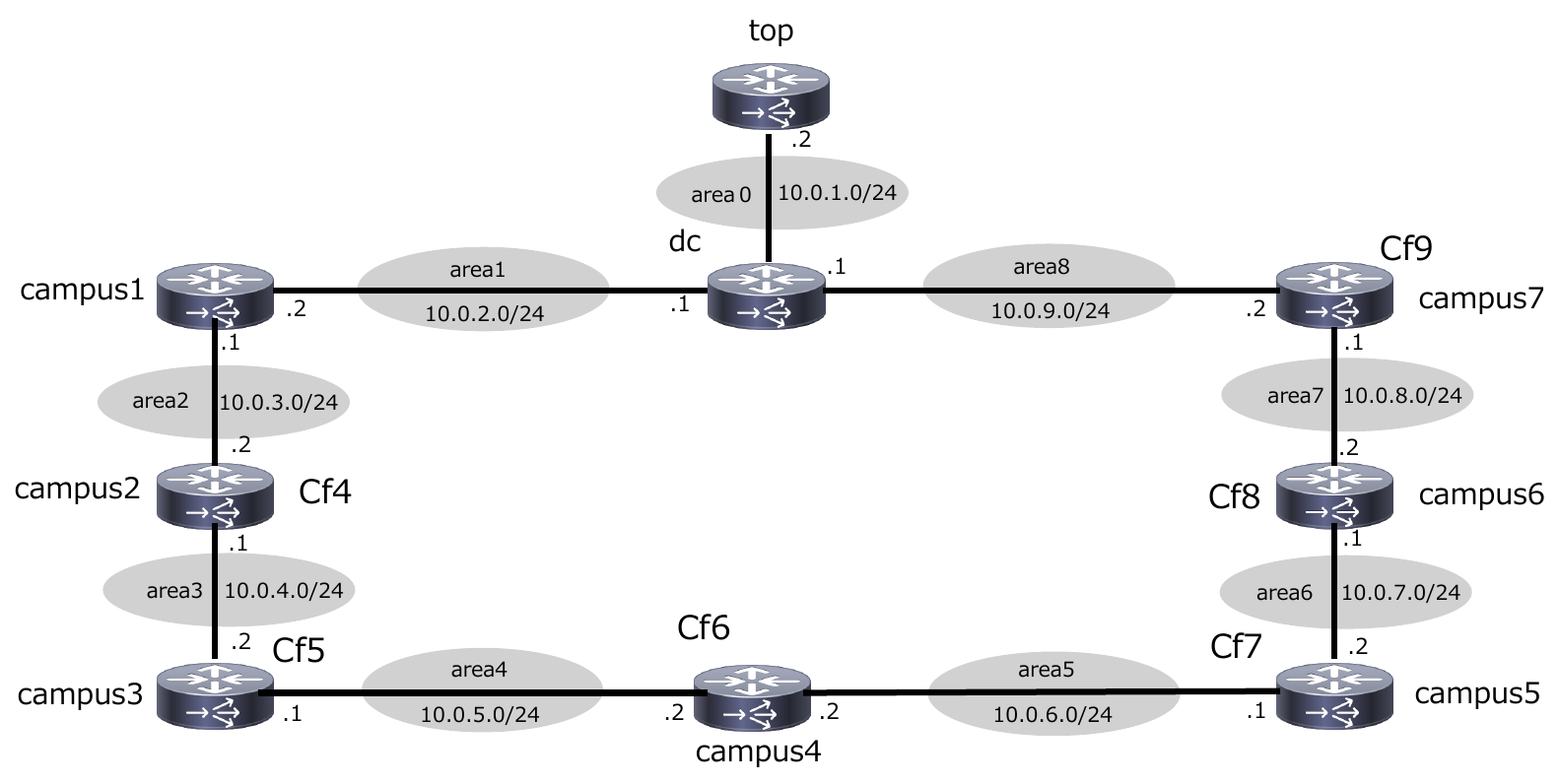}
  \caption{Network used for evaluation}
  \label{fig:setumeijirei}
\end{figure*}

\section{Evaluation}
To demonstrate the effectiveness of the proposed method, we evaluated its ability to detect policy violations in networks containing configuration errors, identify the configuration values responsible for those violations, and generate model transformation outputs equivalent to the actual state information obtained from devices.
In this evaluation, the correctness of both the constructed network configuration model and the resulting network environments was validated by multiple experts in network operations and management.

\subsection{Case Study Network}
\begin{figure*}[!t]
  \centering

  \includegraphics[width=0.73\linewidth]{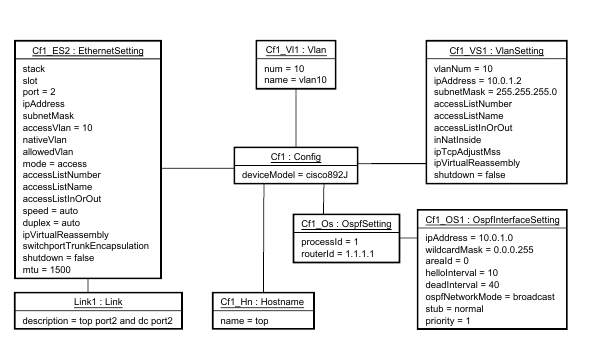}
  
  \caption{Network configuration model for input (top in Fig.4)}
  \label{fig:zu5}
\end{figure*}

This subsection describes the structure of the case study network used in the evaluation.
The network illustrated in Fig.~\ref{fig:setumeijirei} reproduces the configuration deployed at Shinshu University.
Because this study extends our previous work \cite{arai}, we adopted the same case study described in \cite{arai}.
Starting from the correct model provided in the prior study, we constructed an updated correct model by incorporating additional configuration values into the extended model.
A portion of the network configuration model used as input for this experiment is shown in Fig.~\ref{fig:zu5}.
The model shown in Fig.~\ref{fig:zu5} represents the network in Fig.~\ref{fig:setumeijirei} according to the network configuration modeling approach.

In the case study network, a dedicated router is deployed for the network access point (top), the data center (dc), and each of the seven campuses (campus1–campus7), with different VLANs configured between the network devices.
All network devices are required to reach the Internet through the top router, and the connections for this purpose are redundantly configured using a ring topology.
All inter-device links belong to the OSPF domain. The link between the dc and top routers is assigned to Area 0, while different areas are assigned to each segment to enable dynamic routing control.
Additionally, to ensure that every area maintains connectivity to Area 0, VirtualLinks are configured among all network devices except for the direct link between the top and dc routers.

\begin{table*}[t]
\centering
\caption{Overview of Misconfiguration Used in Evaluation and Detection Results}
\label{tab:ayamari}
\renewcommand{\arraystretch}{1.1}
\scalebox{0.80}{
\begin{tabular}{p{2cm}p{5cm}p{9cm}p{1cm}p{3cm}}
\hline\hline
Error setter & Name of the misconfigured item     &  Method to reproduce the misconfiguration    & Result  &Verification items that contributed to detection \\ \hline 
AI& Misconfigured Area ID & Set area of 10.0.5.0 0.0.0.255 in campus3 to 3 (should be Area 4)& $\circ$ & Table 5-(3)\\ \hline
AI & Duplicate Router ID & Set router ID in dc to 1.1.1.1 (duplicate with campus1) & $\circ$ & Table 5-(2) \\ \hline
AI & Missing required network declaration & Removed configuration 'network 10.0.2.0 0.0.0.255 area 1' in campus1 & $\circ$ & Table 5-(9) \\ \hline
AI + Experimenter&Area configured without connection to Area 0 & Removed VirtualLink configuration between campus1 and campus2 & $\circ$ & Table 5-(8)\\ \hline
AI + Experimenter & Insufficient Virtual Links & Removed VirtualLink configuration in campus1 & $\circ$\footnotemark[2] & Table 5-(11) \\ \hline
AI & Timer mismatch & Set hello timer of interfaces in campus2 to 30 seconds (hello timers in campus1 and campus3 are 10 seconds) &     $\circ$ & Table 5-(5) \\ \hline

AI & Missing Area 0 configuration & Set area between dc and top to 9 (should be Area 0) & $\circ$\footnotemark[3] & Table 5-(12) \\ \hline
AI & VLAN mismatch with peer device & Configured different VLANs between campus1 and campus2 (VLAN 30 on campus1, VLAN 40 on campus2) & $\circ$ & Table 4-(5) \\ \hline
AI & Incorrect VLAN configuration on trunk port & Configured different VLANs between campus2 and campus3 (VLANs 1 and 40 on campus2, VLANs 1 and 50 on campus3) & $\circ$ & Table 4-(4) \\ \hline
AI + Experimenter & Missing SVI IP address & Removed 'ip address 10.0.7.1 255.255.255.0' configuration on SVI for VLAN 70 in campus6 & $\circ$ & Table 5-(7), Table 5-(9) \\ \hline
AI & Duplicate IP address & Configured the same IP address (10.0.4.1) on SVIs in campus2 and campus3 & $\circ$ & Table 4-(2) \\ \hline
AI & Incorrect subnet mask configuration & Set subnet mask of SVI for VLAN 20 in campus1 to 255.255.0.0 (should be 255.255.255.0) & $\circ$ & Table 4-(11) \\ \hline
AI&Release of unused interfaces & Enabled unused port (Port 4) in campus3 & $\circ$\footnotemark[2] & Table4-(12) \\ \hline
AI & Duplex/speed mismatch & Set duplex on interfaces between campus4 and campus5 to half on campus4 and full on campus5 & $\circ$ & Table 4-(7) \\ \hline
AI & MTU mismatch & Set MTU on interfaces between campus6 and campus7 to 1500 on campus6 and 1400 on campus7 & $\circ$ & Table 5-(10) \\ \hline
AI & Duplicate subnet usage across different segments & Set subnet of area between campus2 and campus3 to 10.0.2.0/24 (overlaps with Area 1) & $\circ$ & Table 4-(10) \\ \hline
\end{tabular}
}
\vskip 0.3em
*2 Initially in the experiment: ×, after adding verification items: $ \circ $. *3 Initially in the experiment: $ \triangle $, after adding verification items: $\circ$.
\end{table*}
\begin{figure}[t]
\centering
    \begin{lstlisting}[frame=lines, numbers=left]
edges": [
      {
        "node1": {
          "hostName": "top",
          "interfaceName": "FastEthernet2"
        },
        "node2": {
          "hostName": "dc",
          "interfaceName": "FastEthernet2"
        }
      },
  {
        "node1": {
          "hostName": "dc",
          "interfaceName": "FastEthernet4"
        },
        "node2": {
          "hostName": "campus7",
          "interfaceName": "FastEthernet4"
        }
      },
    ...
]


    \end{lstlisting}
    \jecaption{File Summarizing Physical Connections (partial)}{lst_buturi}
    
\end{figure}

\subsection{Experiment 1: Evaluation of Policy Violation Detection and Cause Identification Accuracy}
We evaluated the ability to detect policy violations and to identify the configuration values responsible for those violations in networks with configuration errors.

\subsubsection{Preparation}
The experimental procedure is shown below.
\begin{enumerate}[label=Step \arabic*, leftmargin=3.2em]
\item Sixteen copies of the model were created, and the configuration errors listed in TABLE~\ref{tab:ayamari} were injected into each copy.
\item The proposed method was then applied to each model to obtain the verification results.
\item Based on these results, we examined whether the injected configuration errors and the resulting policy violations were successfully detected.
\end{enumerate}

In this evaluation, we ensured the validity of the experiment by using a generative AI to produce configuration errors that were inserted into the models.
TABLE~\ref{tab:ayamari} provides an overview of the configuration errors used in the evaluation.
In TABLE~\ref{tab:ayamari}, “AI” refers to configuration errors generated directly by the generative AI, whereas “AI+Experimenter” refers to errors refined or selected by the experimenter based on those AI-generated errors.

We describe the method used to derive configuration errors with the assistance of a generative AI.
In this evaluation, we employed ChatGPT to generate configuration errors, using its file-input capability and the latest model available at the time of the experiment (October 2025), GPT-4o.
The procedure and prompts used for generating these configuration errors are presented below.

\begin{enumerate}[label=Step \arabic*, leftmargin=3.2em]
\item Provide the network information to ChatGPT using the prompt: ``The following information reproduces a network operated at a certain university. Since the OSPF protocol is adopted, the operational design considers a redundant configuration with multiple areas. Therefore, in the event of a network failure at any location, traffic will automatically switch to an alternative path. The hostnames of the network devices are campus1 to campus7, dc (data center), and top (external network gateway). First, I will provide the physical connectivity information of this network. Next, I will provide the show running-config information for each router, so please learn from it.''

\item Provide the network's physical connectivity information to ChatGPT in JSON format along with the prompt: ``I am sending the physical connectivity information of the network.''

\item Provide the configuration file of each device to ChatGPT with the prompt: ``I am sending the configuration file of $<$hostname$>$.'' Repeat this for all devices.

\item Input the prompt: ``Please consider and provide possible configuration mistakes for this network. If we were to introduce mistakes into the devices, what mistakes would they be?'' and record the suggested configuration errors. 

\item Input the prompt: ``If there are any other possible mistakes, please include as many as possible.'' to generate additional configuration errors.  

\item Repeat the previous step until ChatGPT no longer outputs configuration errors that differ from those generated previously. 

\end{enumerate}

List \ref{lst_buturi} presents the JSON file representing the physical connectivity information provided in Step 2. In this file, edges is an array representing the physical connections within the network, with node1 and node2 indicating the connected devices. hostName specifies the device hostname, and interfaceName indicates the interface used. In the example shown in List \ref{lst_buturi}, the devices top and dc depicted in Fig.~\ref{fig:setumeijirei} are physically connected via FastEthernet port 2. Moreover, the device configuration files provided in Step 3 were obtained by constructing the network in Fig.~\ref{fig:setumeijirei} using Cisco 1812J and Cisco 892J devices and exporting their configurations using the show running-config command.
Next, the configuration errors output by ChatGPT were selected through consultation with multiple network operations experts. During the selection, only errors in interface settings, VLAN settings, and OSPF-related configurations—those relevant to the proposed method and detectable through static analysis—were included. For example, ChatGPT suggested the following OSPF error: ``a specific area should be configured as a stub area but is incorrectly set as a normal area.'' Since it is not possible to determine from the device configuration alone whether a particular area should be a stub area, this error was therefore excluded from the evaluation.
As an example of a verification item, ``Misconfigured Area ID'' in TABLE~\ref{tab:ayamari} is explained.``Misconfigured Area ID'' refers to a state in which a router is configured with an erroneous area ID. As a result, adjacent devices have mismatched area IDs, preventing the formation of neighbor relationships and causing routing information to be exchanged incorrectly. In this experiment, the error was reproduced by misconfiguring campus3’s OSPF-enabled area: instead of 10.0.5.0 0.0.0.255 area 4, it was set as 10.0.5.0 0.0.0.255 area 3, and this configuration was added to the model.

\subsubsection{Results}

The experimental results and the verification items that contributed to detecting configuration errors are presented in TABLE~\ref{tab:ayamari}. Verification items that contributed to error detection indicate which items were effective in identifying specific errors. For example, Table 5-(3) for`` Misconfigured Area ID'' corresponds to the detection performed by Table 5, item (3)`` Area ID mismatch.''
In TABLE~\ref{tab:ayamari}, a circle ($\circ$) indicates that both the policy violation and the configuration value causing it were correctly detected and identified, a triangle ( $\triangle$ ) indicates that the policy violation was detected but the cause could not be determined, and a cross (×) indicates that the policy violation was not detected.
For instance, ``Incorrect area ID'' was output as ``Area IDs of campus3\_OS4 and campus4\_OS4 do not match,'' and the corresponding OspfInterfaceSetting node values were correctly identified. Among the 16 configuration errors used in the evaluation, 14 errors (approximately 88\%) were detected as policy violations, and the configuration values causing 13 errors (approximately 81\%) were correctly identified.
--
\subsubsection{Discussion}
For the case of ``Missing Area 0 configuration,'' the verification results indicated that multiple areas (Areas 2–7) were not connected to Area 0, but the absence of Area 0 itself could not be pointed out. This is because the proposed method defined a verification item as ``all areas are connected to Area 0,'' and no item was defined to check ``the existence of Area 0.''
Furthermore, ``Insufficient Virtual Links'' and ``Release of unused interfaces,'' no verification items had been defined. After improving the proposed method and defining these items, all 16 configuration errors used in the evaluation were detected as policy violations (100\%), and the configuration values causing all 16 errors were correctly identified.
Among the 16 configuration errors used in the evaluation, approximately 90\% of policy violations were detected, and about 80\% of the configuration values causing the violations were correctly identified. After improving the proposed method, all policy violations were detected and their causes were correctly identified, suggesting the effectiveness of the proposed method.
An analysis of the contribution of verification items to error detection showed that, among the 27 proposed verification items, 13 items (approximately 48\%) contributed to detecting the configuration errors in this experiment. The remaining 14 items need to be evaluated in future experiments. These items include consistency checks that may arise during model creation (e.g., Missing num in Vlan) and items targeting loops that do not cause issues in OSPF, which may be evaluated in other cases or participant experiments.
In addition, configuration errors selected in consultation with multiple network operations management experts confirmed that the errors output by ChatGPT included many that could occur in real network designs. In particular, ``Incorrect subnet mask configuration'' was found to be a subtle error, since communication may still work even if the setting is wrong. This suggests that ChatGPT can generate such hard-to-notice configuration errors.

\begin{figure*}[t]
    \centering
    \begin{lstlisting}[frame=lines, numbers=left]
Vlan20
  Internet Address 10.0.2.2/24 Area1
  Process ID 1, Router ID 3.3.3.3, Network Type BROADCAST
                          State DR, Priority1
  Designated Router (ID) 3.3.3.3, Interface address 10.0.2.2
  Backup Designated router (ID) 2.2.2.2, Interface address 10.0.2.1
  Timer intervals configured, Hello 10, Dead 40
    \end{lstlisting}
    \jecaption{OSPF information for VLAN 20 in campus1 in model (partial)}{lst_turu}
    
\end{figure*}

\begin{figure*}[t]
    \centering
    \begin{lstlisting}[frame=lines, numbers=left]
campus1#show ip ospf interface vlan 20
Vlan20 is up, line protocol is up
  Internet Address 10.0.2.2/24, Area 1
  Process ID 1, Router ID 3.3.3.3, Network Type BROADCAST, Cost: 1
  Transmit Delay is 1 sec, State BDR, Priority 1
  Designated Router (ID) 2.2.2.2, Interface address 10.0.2.1
  Backup Designated router (ID) 3.3.3.3, Interface address 10.0.2.2
  Timer intervals configured, Hello 10, Dead 40, Wait 40, Retransmit 5 Retransmit 5
    \end{lstlisting}
    
    \jecaption{OSPF information for VLAN 20 in campus1 in the physical network (partial)}{lst_jikki}
    
\end{figure*}

\subsection{Experiment 2: Evaluation of Design Information Output via Model Transformation}

This experiment evaluated whether the design information produced by the model transformation is equivalent to the status information obtained from the actual network devices.

\subsubsection{Preparation}

In this experiment, the model corresponding to the case study was input into the proposed tool, and the obtained information was compared with the information acquired by executing status-check commands on a network constructed with actual devices.

The status-check commands used for the comparison were as follows:
\begin{enumerate}
\item show running-config
\item show vlan brief
\item show ip protocols
\item show ip ospf interface
\item show ip ospf neighbor
\end{enumerate}
To construct the actual network, Cisco 892J devices (top, dc) and Cisco 1812J devices (campus1–campus7) were employed.
\subsubsection{Results}

The OSPF operational status of each interface as obtained by the proposed tool is shown in List \ref{lst_turu}, while the OSPF operational status obtained from the actual devices is shown in List \ref{lst_jikki}.
When comparing the outputs from the five status-check commands, all information could be produced in a format that closely resembles that of the actual devices, except for the devices selected as DR and BDR.

\subsubsection{Discussion}

The difference observed in the devices selected as DR and BDR can be attributed to the fact that the proposed method selects DR and BDR based on the OSPF priority values and router IDs of each network device, regardless of the order in which devices are 
In this experiment, the routers selected as DR and BDR varied depending on the sequence in which configurations were applied to the devices, leading to variations in the outputs.
Although such differences were observed, all other information was confirmed to be equivalent. Therefore, the output of design information via model transformation using the proposed method is effective as a means for designers to verify network device information.

\section{Related Work}

This section reviews previous studies that perform verification based on network device configuration files and network topology, and discusses how they differ from the proposed method.

Tajima et al. \cite{kijokensa_tajima} proposed a method that enables verification of whether design policies are correctly reflected without using actual devices. However, their work does not evaluate the capability to identify the specific configuration values that cause policy violations. In contrast, the proposed method demonstrates through the evaluation that when a policy violation is detected, it can pinpoint the incorrect configuration values.

Batfish \cite{batfish} transforms network device configuration files and supplementary network information into a declarative language called Datalog, generating a data plane. It automatically checks multipath consistency, fault-tolerance consistency, and destination reachability using the generated data-plane relations. In contrast, the proposed method offers an advantage in verifying additional types of consistency, such as VLAN segment duplication and duplex mismatches.

RCC \cite{rcc} defines accuracy conditions concerning path visibility and route validity, such as duplicate router IDs or configurations that lead to unreachable BGP destinations, and automatically identifies violations. In contrast, the proposed method performs cross-protocol consistency verification and identifies policy violations.

A common feature of these related works is their reliance on network device configuration files as input. These configuration files are originally obtained from actual devices, and applying these methods to newly constructed networks requires manually creating corresponding configuration files. However, these approaches do not handle verification for syntactically incomplete configuration files. In contrast, the proposed tool uses a network configuration model that can serve as a design specification and performs verification on models even when they contain syntactically incomplete configuration values.

\section{Conclusion}

In this paper, we proposed an automatic verification method based on static analysis targeting network configuration model. The proposed method accepts a network configuration model as input and conducts consistency verification to detect policy violations and identify the configuration values responsible. We used an operational network as a case study, intentionally introducing configuration errors to evaluate the effectiveness of the proposed method. The results indicated that approximately 88\% of policy violations were detected. After refining the proposed method, all policy violations and their root causes were successfully identified, confirming the method’s effectiveness.

Future work includes expanding the set of verification items and assessing their effectiveness. For example, the current selection of OSPF DR and BDR routers assumes router ID settings, so no verification has been conducted that considers loopback addresses. Similarly, the OSPF authentication feature has not been verified in terms of consistency with neighboring devices based on authentication presence or method. Furthermore, the detection of IP address duplication is not yet complete. For instance, the ipAddress setting of IpRoute indicates the destination network for static routing, but static routing configurations were not included in the verification scope; therefore, duplication of IpRoute ipAddress values was not checked. These issues partly stem from the limited expressiveness of the meta-model, and addressing them requires extending the metamodel and implementing corresponding verification
Another potential future direction is verifying devices from other vendors. In this paper, experiments were conducted using Cisco 1812J and Cisco 892J. Since the specification items of the network configuration metamodel reference command structures common to Cisco devices\cite{arai}, the method can be applied not only to Cisco 1812J but also to a wide range of models. However, the current meta-model primarily targets Cisco devices. For non-Cisco devices, the proposed method could potentially be applied by mapping device-specific configuration items to the network configuration model and assigning corresponding values, but this mapping must be performed manually, and any items not representable in the meta-model would require extension. Therefore, research is underway to expand the network configuration model to support a broader range of devices\cite{nakamura}. In the future, the verification scope of the proposed method will be expanded according to the extended model, and its effectiveness will then be evaluated.

Furthermore, as part of our efforts toward network automation, we have proposed automatic verification methods using simulators without actual devices\cite{satake} and automatic generation of device configuration procedures\cite{arai}, both targeting network configuration model. While the simulator-based method can verify fault scenarios not addressed by the proposed method, the proposed method statically executes commands to check device configurations, which imposes certain limitations. For example, dynamic information obtained during device operation, such as routing tables from show ip route, cannot be verified. In the future, combining dynamic verification methods based on network configuration model\cite{satake} with the proposed method is expected to overcome these limitations, and the effectiveness of the improved approach will be evaluated.

\bibliographystyle{IEEEtran}
\bibliography{mybib}

@techreport{downtime,
    author ={{Juniper Networks}},
    title = {Whats Behind Network Downtime? Proactive Steps to Reduce Human Error and Improve Availability of Networks},
    institution = {Juniper Networks},
    year = {2008}
}

@article{arai,
   title={A System to Automatically Generate Configuration Instructions for Network Elements from Network Configuration Models},
   author={Arai, Nagi and Ogata, Shinpei and Suzuki, Hikofumi and Okano, Kozo},
   journal={Journal of digital practices},
   year={2023},
   volume={4},
   number={3},
   pages={33--47},
}

@inproceedings{gember2016fast,
  title={Fast control plane analysis using an abstract representation},
  author={Gember-Jacobson, Aaron and Viswanathan, Raajay and Akella, Aditya and Mahajan, Ratul},
  booktitle={Proceedings of the 2016 ACM SIGCOMM Conference},
  pages={300--313},
  year={2016}
}

@article{arai1,
  title={Model-Driven Approach for Supporting
Configuration of Network Devices},
  author={Arai, Nagi and Ogata, Shinpei and Suzuki, Hikofumi and Okano, Kozo },
  journal={IPSJ SIG Technical Reports},
  volume={2023},
  number={15},
  pages={1--8},
  year={2023}
}

@book{miyata_siken,
    author = {Hiroshi Miyata},
    title = {Introduction to Network Functional Testing for Infrastructure and Network Engineers},
    publisher = {SB Creative Corp.},
    year = {2021}
}

@inproceedings{rcc,
  title={Detecting BGP configuration faults with static analysis},
  author={Feamster, Nick and Balakrishnan, Hari},
  booktitle={Proceedings of the 2nd conference on Symposium on Networked Systems Design \& Implementation-Volume 2},
  pages={43--56},
  year={2005}
}

@article{kijokensa_tajima,
  title={Support System of Designing and Verifying Network by Abstracting Topology Model from Configuration Files},
  author={Tajima, Teruhisa and Kawaguchi, Eiichiro and Takiguchi, Toshiyuki and Hagiwara, Manabu and Shinzato, Yasuaki},
  journal={IEICE Technical Report},
  volume={122},
  number={96},
  pages={54--59},
  year={2022},
}

@article{nakamura,
   title={Automatic Extraction of Network Configuration Model by Analyzing Configuration of Network Elements -Toward Multi-Vendor Support-},
  author={Nakamura, Kosei and Suzuki, Hikofumi and Ogata, Shinpei and Hashiura, Hiroaki and Okano, Kozo},
  journal={IPSJ SIG Technical Reports},
  volume={2024},
  number={6},
  pages={1--8},
  year={2024}
}

@inproceedings{batfish,
author = {Fogel, Ari and Fung, Stanley and Pedrosa, Luis and Walraed-Sullivan, Meg and Govindan, Ramesh and Mahajan, Ratul and Millstein, Todd},
title = {A General Approach to Network Configuration Analysis},
year = {2015},
isbn = {9781931971218},
publisher = {USENIX Association},
address = {USA},
booktitle = {Proceedings of the 12th USENIX Conference on Networked Systems Design and Implementation},
pages = {469–483},
numpages = {15},
location = {Oakland, CA},
series = {NSDI'15}
}

@article{fujita_iot,
  title={Automatic Verification of Network Configuration Model by Static Analysis},
  author={Fujita, Tomoya and Suzuki, Hikofumi and Ogata, Shinpei and Hashiura, Hiroaki and Okano, Kozo},
  journal={IPSJ SIG Technical Reports},
  volume={2024},
  number={5},
  pages={1--8},
  year={2024}
}

@article{satake,
  title={Proposal and Evaluation of a Tool for Validating Link-Failure Behavior of Network in Design},
  author={Satake, Shuji and Suzuki, Hikofumi and Ogata, Shinpei and Okano, Kozo},
  journal={Journal for Academic Computing and Networking},
  volume={27},
  number={1},
  pages={180--190},
  year={2023},
}

\end{document}